\documentclass[twocolumn,english,aps,prd,graphics,floatfix,a4paper,superscriptaddress]{revtex4-1} % {revtex4}
\usepackage{amsmath, amssymb}
\usepackage{makecell}
\usepackage{multirow}
\usepackage{CJK}
\usepackage{graphicx}
\usepackage{mathrsfs}
\usepackage{bm}
\usepackage{amsmath}
\usepackage{dcolumn}
\usepackage{epstopdf}
\usepackage{dsfont}
\usepackage{amssymb}
\usepackage{tabularx}
\usepackage{array}
\usepackage{float}
\usepackage{color}
\usepackage{epstopdf}
\usepackage{mathrsfs}
\usepackage[colorlinks, linkcolor=blue,anchorcolor=blue,citecolor=blue,urlcolor=blue]{hyperref}

\begin{document}
%\begin{CJK*}{GBK}{song}

\title{First-principles study on the bulk and two-dimensional structures of $A$MnBi($A=$K, Rb, Cs)-family materials}

\author{Ziming Zhu}\email{zimingzhu@hunnu.edu.cn}
\affiliation{Key Laboratory of Low-Dimensional Quantum Structures and Quantum Control of Ministry of Education, Department of Physics and Synergetic Innovation Center for Quantum Effects and Applications, Hunan Normal University, Changsha 410081, China}

\author{Chunyan Liao}
\affiliation{Key Laboratory of Low-Dimensional Quantum Structures and Quantum Control of Ministry of Education, Department of Physics and Synergetic Innovation Center for Quantum Effects and Applications, Hunan Normal University, Changsha 410081, China}

\author{Si Li}
\affiliation{Key Laboratory of Low-Dimensional Quantum Structures and Quantum Control of Ministry of Education, Department of Physics and Synergetic Innovation Center for Quantum Effects and Applications, Hunan Normal University, Changsha 410081, China}
\affiliation{Research Laboratory for Quantum Materials, Singapore University of Technology and Design, Singapore 487372, Singapore}

\author{Xiaoming Zhang}
\affiliation{School of Materials Science and Engineering, Hebei University of Technology, Tianjin 300130, China}

\author{Weikang Wu}
\affiliation{Research Laboratory for Quantum Materials, Singapore University of Technology and Design, Singapore 487372, Singapore}

\author{Zhi-Ming Yu}
\affiliation{Key Lab of Advanced Optoelectronic Quantum Architecture and Measurement (MOE), Beijing Key Lab of Nanophotonics \& Ultrafine
Optoelectronic Systems, and School of Physics, Beijing Institute of Technology, Beijing 100081, China}

\author{Rui Yu}
\affiliation{School of Physics and Technology, Wuhan University, Wuhan 430072, China}

\author{Wei Zhang}\email{zhangw721@163.com}
\affiliation{Fujian Provincial Key Laboratory of Quantum Manipulation and New Energy Materials, College of Physics and Energy, Fujian Normal University, Fuzhou 350117, China}

\author{Shengyuan A. Yang}
\affiliation{Research Laboratory for Quantum Materials, Singapore University of Technology and Design, Singapore 487372, Singapore}

\begin{abstract}
Magnetic materials with high mobilities are intriguing subject of research from both fundamental and application perspectives. Based on first-principle calculations, we investigate the physical properties of the already synthesized $A$MnBi($A=$K, Rb, Cs)-family materials. We show that these materials are antiferromagnetic (AFM), with Neel temperatures above 300 K. They contain AFM ordered Mn layers, while the interlayer coupling changes from ferromagnetic (FM) for KMnBi to AFM for RbMnBi and CsMnBi. We find that these materials are narrow gap semiconductors. Owing to the small effective mass, the electron carrier mobility can be very high, reaching up to $10^5$ cm$^2$/(V$\cdot$s) for KMnBi. In contrast, the hole mobility is much suppressed, typically lower by two orders of magnitude. We further study their two-dimensional (2D) single layer structures, which are found be AFM with fairly high mobility $\sim 10^3$ cm$^2$/(V$\cdot$s). Their Neel temperatures can still reach room temperature.  Interesting, we find that the magnetic phase transition is also accompanied by a metal-insulator phase transition, with the paramagnetic metal phase possessing a pair of nonsymmorphic-symmetry-protected 2D spin-orbit Dirac points. Furthermore, the magnetism can be effectively controlled by the applied strain. When the magnetic ordering is turned into FM, the system can become a quantum anomalous Hall insulator with gapless chiral edge states.

\end{abstract}

\maketitle
%\end{CJK*}
\section{Introduction}

Layered magnetic materials have received great attention in recent years. A major driving force is their potential for achieving two-dimensional (2D) magnetism. Indeed, the first few examples of 2D magnetic materials, such as CrI$_3$~\cite{HuangLayer}, Cr$_2$Ge$_2$Te$_6$~\cite{GongDiscovery}, and Fe$_3$GeTe$_2$~\cite{Zhu2015Electronic,Zhuang2016Strong,DengGate}, have been obtained by exfoliation from their 3D layered counterparts. So far, there are two challenges in the field of 2D magnetic materials. First, the magnetism in the discovered examples is still weak, with low magnetic transition temperatures (e.g., 45 K for CrI$_3$~\cite{HuangLayer} and 61 K for Cr$_2$Ge$_2$Te$_6$~\cite{GongDiscovery}). This severely hinders the experimental study as well as practical applications. Second, the concept of antiferromagnetic (AFM) spintronics has been demonstrated in recent works~\cite{WadleyElectrical,JungwirthAntiferromagnetic}. Compared to ferromagnets (FMs), AFM materials have the advantages of no stray fields and relative insensitivity to external magnetic fields, which are desired for small devices. However, the progress on the 2D AFM materials is still slow. Experimentally, intrinsic 2D AFM has only been reported in FePS$_3$ nanosheets, with a low Neel temperature $\sim$118 K~\cite{Lee2017Ising,XingzhiRaman}. To address these challenges, a natural idea is to explore existing 3D layered materials with strong intralayer AFM ordering, and to realize 2D AFM layers based on them.

In this work, we adopt this idea and investigate the family of $A$MnBi ($A=$K, Rb, Cs) materials. These materials have been synthesized in experiment~\cite{LompwskySynthese}. They all share a layered tetragonal crystal structure, with AFM ordering at low temperature determined in the {neutron diffraction} experiment~\cite{Schucht1999ChemInform}. However, many important physical properties, such as the magnetic transition temperatures and the electronic properties, remain unexplored for these materials. Here, by using first-principles calculations, we perform a systematic study on their physical properties. We find that these materials all have robust AFM ordering within each Mn atomic layer. The interlayer coupling is of FM type for KMnBi, whereas it is of AFM type for RbMnBi and CsMnBi, which are consistent with the previous experimental observation~\cite{Schucht1999ChemInform}. Using Monte-Carlo simulations, their Neel temperatures are estimated to be around the room temperature. In the ground state, these materials are narrow-gap semiconductors with band gaps about {0.3 eV}.  We find that the electron and hole carriers have distinct properties. While the effective mass is small for electron carriers; the hole effective mass, especially along the out-of-plane direction, is much larger. As a consequence, we find that the electron mobility can be very high, even reaching $10^5$ cm$^2$/(V$\cdot$s) for KMnBi. In comparison, the hole mobility is lower by about two orders of magnitude.
We then investigate the 2D single-layer structures of these materials. We show that these 2D layers are stable and remain AFM in the ground state. Their Neel temperatures can still reach the room temperature, and they maintain a relatively high mobility $\sim 10^3$ cm$^2$/(V$\cdot$s).
Interestingly, the magnetic phase transition is simultaneously a metal-insulator phase transition, with the paramagnetic phase being a metal with protected 2D spin-orbit Dirac points. Further, we show that
the magnetism in these 2D layers can be effectively controlled by the applied strain. For 2D KMnBi, its Neel temperature can be decreased by more than 250 K by an applied 5\% strain. In addition, if the magnetic ordering is turned into FM, e.g., by coupling with a FM substrate, the system would become a quantum anomalous Hall insulator with gapless chiral edge states. Our work reveals interesting physics of an existing family of magnetic materials, and provides a promising platform to explore 2D materials with robust magnetism and high mobility.

\section{Computational Method}

Our first-principle calculations were based on the density functional theory (DFT), using a plane-wave basis set and projector augmented wave method~\cite{PhysRevB.50.17953}, as implemented in the Vienna \emph{ab} \emph{initio} simulation package (VASP)~\cite{PhysRevB.54.11169,kresse1999ultrasoft}. The generalized gradient
approximation (GGA) parameterized by Perdew, Burke, and Ernzerhof (PBE) was adopted for the exchange-correlation functional.
The energy cutoff was set to 360 eV, and a $11\times11\times5$ Monkhorst-Pack $k$ mesh was used for the Brillouin zone (BZ) sampling. The lattice constants for the bulk calculations were fixed to the experimental values (shown in Table~\ref{table:bulk}). The atomic positions were fully optimized until the residual forces were less than $10^{-3}$ eV/{\AA}. The convergence criterion for the total energy was set to be $10^{-8}$ eV. To account for the correlation effects on the transition metal Mn, the GGA$+U$ method with $U=5$ eV for Mn-3\emph{d} orbitals was adopted~\cite{otrokov2019unique}.

\begin{table}
	\centering
	\caption{Lattice constants (experimental value, in \AA) and the total free energy per formula unit (in meV) for the different magnetic configurations. Here, the ground state magnetic configuration is taken as the reference (with its energy set as zero).}
	\label{my-label}
	\renewcommand\arraystretch{1.4}
	\begin{tabular}{p{1.25cm}<{\centering}p{0.75cm}<{\centering}p{0.75cm}<{\centering}p{1.25cm}<{\centering}p{1.25cm}<{\centering}p{1.25cm}<{\centering}p{1.25cm}<{\centering}}
		\hline\hline
		\rule{0pt}{13pt}
		&$a$   &$c$  & FM  & A-type & C-type  & G-type \\
		\hline
		KMnBi  &$4.723$  &$8.269$  &$78.5$    &$102$  & $0$  &$0.178$\\
		RbMnBi &$4.785$   &$8.542$  &$76.4$   &$125$   &$0.274$  &$0$ \\
		CsMnBi     &$4.843$   &$8.933$   &$73.5$   & $126$ &$0.001$  &$0$\\
		\hline\hline
	\end{tabular}
	\renewcommand\arraystretch{1.4}
	\label{table:bulk}
\end{table}

The Phonon spectra were calculated by using the PHONONPY code~\cite{Togo2008First}, within the density functional perturbation theory~\cite{Gonze1997Dynamical}. The magnetic transition temperatures were estimated by Monte-Carlo simulations using the VAMPIRE package~\cite{EvansAtomistic}. In the simulation, we used a supercell with size $15\times 15\times 15$.

To study the topological properties and surface states, we constructed an \emph{ab initio} tight binding model based on the Wannier functions~\cite{marzari1997maximally,PhysRevB.65.035109,RevModPhys.84.1419}. The surface spectra were investigated by using this model and by the iterative Green's function method~\cite{sancho1985highly}, as implemented in the WannierTools package~\cite{wu2017wanniertools}. The Chern number was calculated via the Wannier charge center method~\cite{PhysRevB.95.075146}.

\section{3D Bulk Property}

We first investigate the properties of the 3D bulk phase of the $A$MnBi ($A=$K, Rb, Cs) family materials.

\subsection{Crystal Structure}

\begin{figure}[!t]
{\includegraphics[clip,width=8.2cm]{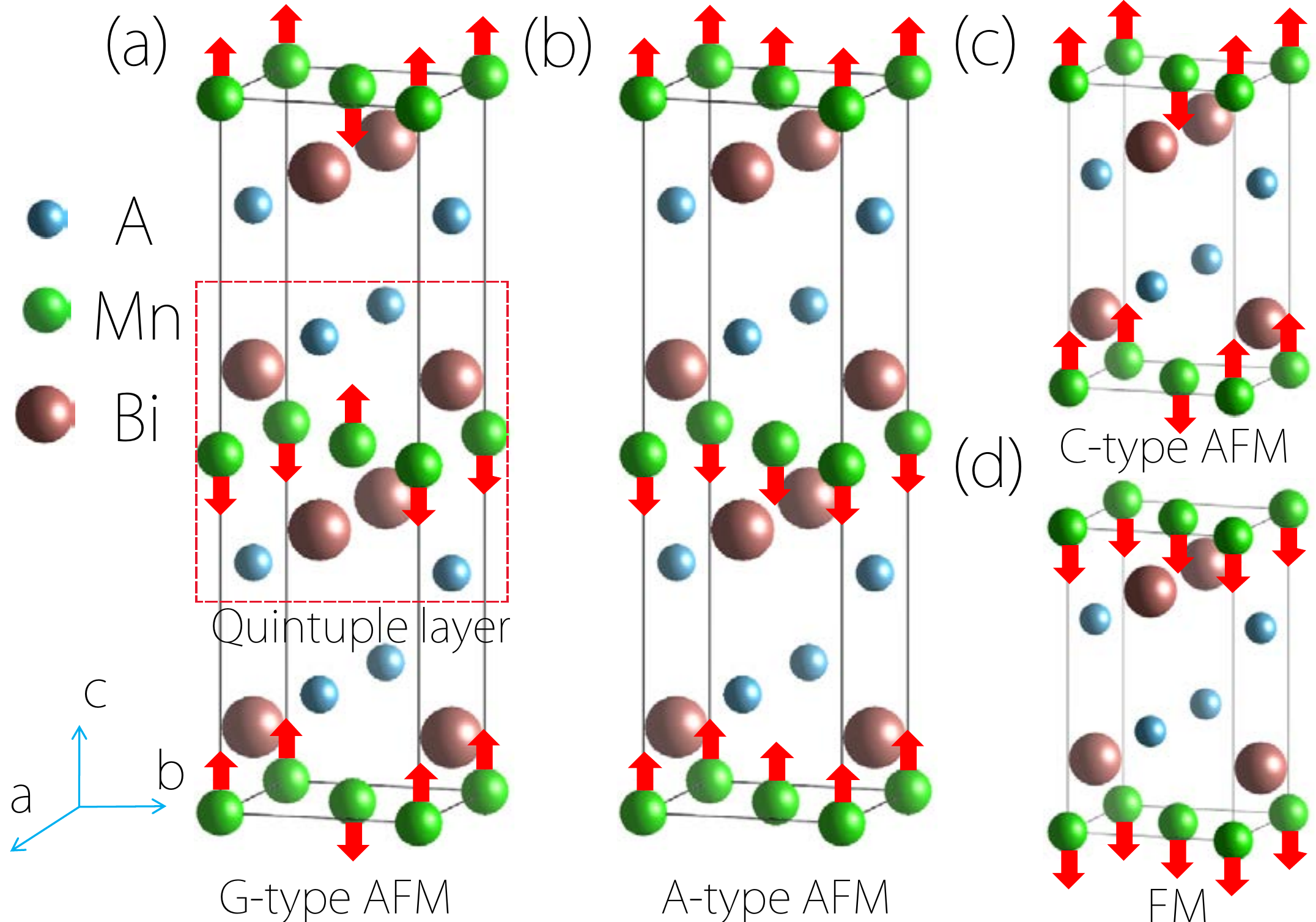}}
%magstru_3.png
\caption{\label{fig:ms} Crystal structure of $A$MnBi and the four considered magnetic configurations: (a) G-type AFM, (b) A-type AFM, (c) C-type AFM, and (d) FM. }
\end{figure}

The $A$MnBi ($A=$K, Rb, Cs) compounds are a family of alkali metal manganese pnictides. They were first reported in experiment in  1974~\cite{LompwskySynthese}, obtained from the elements by high temperature reactions. In single crystals, these compounds share the same kind of tetragonal crystal structure,
with space group $D^{7}_{4h}$ (No.~129) with 6 atoms (i.e., two formula units) in a primitive unit cell. As shown in Fig.~\ref{fig:ms}, these materials have a layered form: each unit corresponds to a quintuple layer unit, consisting of five atomic layers ordered in the $A$-Bi-Mn-Bi-$A$ sequence. Within each atomic layer, the atoms form a 2D square lattice. (The Mn layer has an atomic density which doubles the other atomic layers.) As we will see later, the low-energy bands are mostly from the Bi-$6p$ and Mn-$3d$ orbitals, i.e., the electronic properties are mainly determined by the Bi-Mn-Bi trilayers. Meanwhile, the alkali atoms play a minor role: they act as charge donors and can be regarded as being intercalated into the spacing between the Bi-Mn-Bi layers. The lattice constants for these materials are listed in Table~\ref{table:bulk}. The detailed atomic positions obtained from our calculations are presented in the Supplemental Material~\cite{supp_material}.

The crystal structure possesses the following symmetry generators (besides the translations): the four-fold roto-inversion $\overline{4}_{z}: (x,y,z)\rightarrow(y,-x,-z)$, the horizonal glide mirror $G_z:\,(x,y,z)\rightarrow (x+1/2,y+1/2,-z)$, and the inversion symmetry $\mathcal{P}$. Without magnetic ordering, these crystal symmetries as well as the time reversal symmetry ($\mathcal{T}$) are preserved. In the presence of magnetic ordering, $\mathcal{T}$ and some of the above symmetries would be broken. However, depending on the ordering, certain magnetic symmetry, e.g., the combination of $\mathcal{T}$ and $\mathcal{P}$, may still be preserved, as we will mention below.

\subsection{Magnetic Ordering}

Compounds containing transition metal elements often exhibit magnetic ordering in their ground state. In $A$MnBi, the magnetism mainly comes from Mn. The Mn ions have a nominal valence of $+2$, with five electrons in the $d$ shell. As shown in Fig.~\ref{fig:ms}, each Mn ion is sitting in a tetrahedron formed by the nearby Bi ions. The $3d$ orbitals are split by the tetrahedral crystal field into the $e_g$ and $t_{2g}$ orbitals, with $t_{2g}$ higher in energy. With five $d$ electrons, Mn$^{2+}$ usually take the high spin state, with the configuration of $(e_g^2 t_{2g}^3)$. This leads to a spin of $5/2$ for a single Mn$^{2+}$ ion. This spin magnitude agrees well with the magnetic moment $\sim 4.5\mu_B$ for each Mn site, obtained from DFT calculations (see Table~\ref{table:mag}).

To determine the ground state magnetic structure, we compare the energies of several typical types of magnetic ordering. As illustrated in Fig.~\ref{fig:ms}, these include the FM, the A-type AFM, the C-type AFM, and the G-type AFM. For each type, we also determine the preferred orientation of the spin [with spin-orbit coupling (SOC) included in the calculation]. We find that for all the types, the magnetic easy axis is along the $z$ direction. The calculated energies for the different orderings (with spins in the $z$ direction) are presented in Table~\ref{table:bulk}.

From the results, one observes that, first, in the ground state, the 2D Mn layers in all these materials prefer the AFM ordering. This can be readily understood. In the Mn layer, the Mn ions are close to each other (with bond length $\sim$ $3.34$ \AA), and there are direct overlap between the occupied $d$-orbitals from the neighboring Mn ions. As we shall see, the ground states of these materials are insulators. Hence, the AFM ordering is preferred due to the super-exchange mechanism, and is consistent with the Goodenough-Kanamori-Anderson rules~\cite{Anderson1950Antiferromagnetism,Goodenough1955Theory,GoodenoughAn,KanamoriSuperexchange}. Second, there is an interesting difference when these Mn layers are stacked together inside these materials. The interlayer magnetic coupling is relatively weak, as manifested by the small energy difference between C-type and G-type configurations.
One observes that KMnBi has C-type AFM in the ground state, whereas RbMnBi and CsMnBi prefer G-type AFM in the ground state. In other words, the interlayer coupling in KMnBi is FM, whereas in RbMnBi and CsMnBi is AFM. These results are consistent with a previous neutron diffraction experiment~\cite{Schucht1999ChemInform}.
%However, currently, we do not have a simple reason to explain this difference (the interlayer distance clearly plays an important role here).

We have estimated the magnetic transition temperatures by using the Monte-Carlo simulations based on a classical Heisenberg-like spin model~\cite{EvansAtomistic}:
\begin{equation}\label{Hmodel}
 H=-\sum_{i,j}J_{ij}\bm S^i\cdot \bm S^j-K\sum_i(S_z^i)^2.
\end{equation}
Here, the spin vectors are normalized, $i$ and $j$ label the Mn sites, $J_{ij}$ is the exchange coupling strength between sites $i$ and $j$, and $K$ represents the magnetic anisotropy strength. In the exchange term, we include the first- and second-neighboring intralayer coupling, as well as the first-neighbor interlayer coupling. The values of the parameters $J_{ij}$ and $K$ are extracted from the DFT calculations (see Table~\ref{table:mag}). The result shows that, in these materials, the intralayer coupling are much stronger than the interlayer coupling. The Neel temperatures obtained from the simulations are listed in Table~\ref{table:mag}. One can see that the transition temperatures are all above 300 K, indicating that the AFM orderings in these materials are fairly robust.

\begin{table}
	\centering
	\caption{Calculated magnetic moment $\mu$ per Mn (in $\mu_B$), the parameters in the spin model (\ref{Hmodel}) (in meV), and the estimated Neel temperature $T_\text{N}$ (in K). Here, $J_1$ and $J_2$ are the intralayer first and second neighbor exchange coupling strength, and $J_3$ is the interlayer first neighbor exchange coupling strength.}
	\label{my-label}
	\renewcommand\arraystretch{1.4}
	\begin{tabular}{p{1.0cm}<{\centering}p{1.0cm}<{\centering}p{1.0cm}<{\centering}p{1.25cm}<{\centering}p{1.25cm}<{\centering}p{1.25cm}<{\centering}p{1.25cm}<{\centering}}
		\hline\hline
		\rule{0pt}{13pt}
		&$\mu$   &$J_1$  &$J_2$  &$J_3$  &$K$  &$T_\text{N}$ \\
		\hline
		KMnBi  &$4.49$  &$-196.64$  &$16.79$    &$0.36$  & $1.28$  &$305$\\
		RbMnBi &$4.54$   &$-191.33$  &$17.55$   &$-0.35$   &$1.44$  &$308$ \\
		CsMnBi     &$4.49$   &$-191.28$   &$17.54$   & $-0.35$ &$1.34$  &$306$\\
		\hline\hline
	\end{tabular}
	\renewcommand\arraystretch{1.4}
	\label{table:mag}
\end{table}

\label{3D_Bulk}

\subsection{Electronic Property}

\begin{figure}[!t]
	{\includegraphics[clip,width=8.2cm]{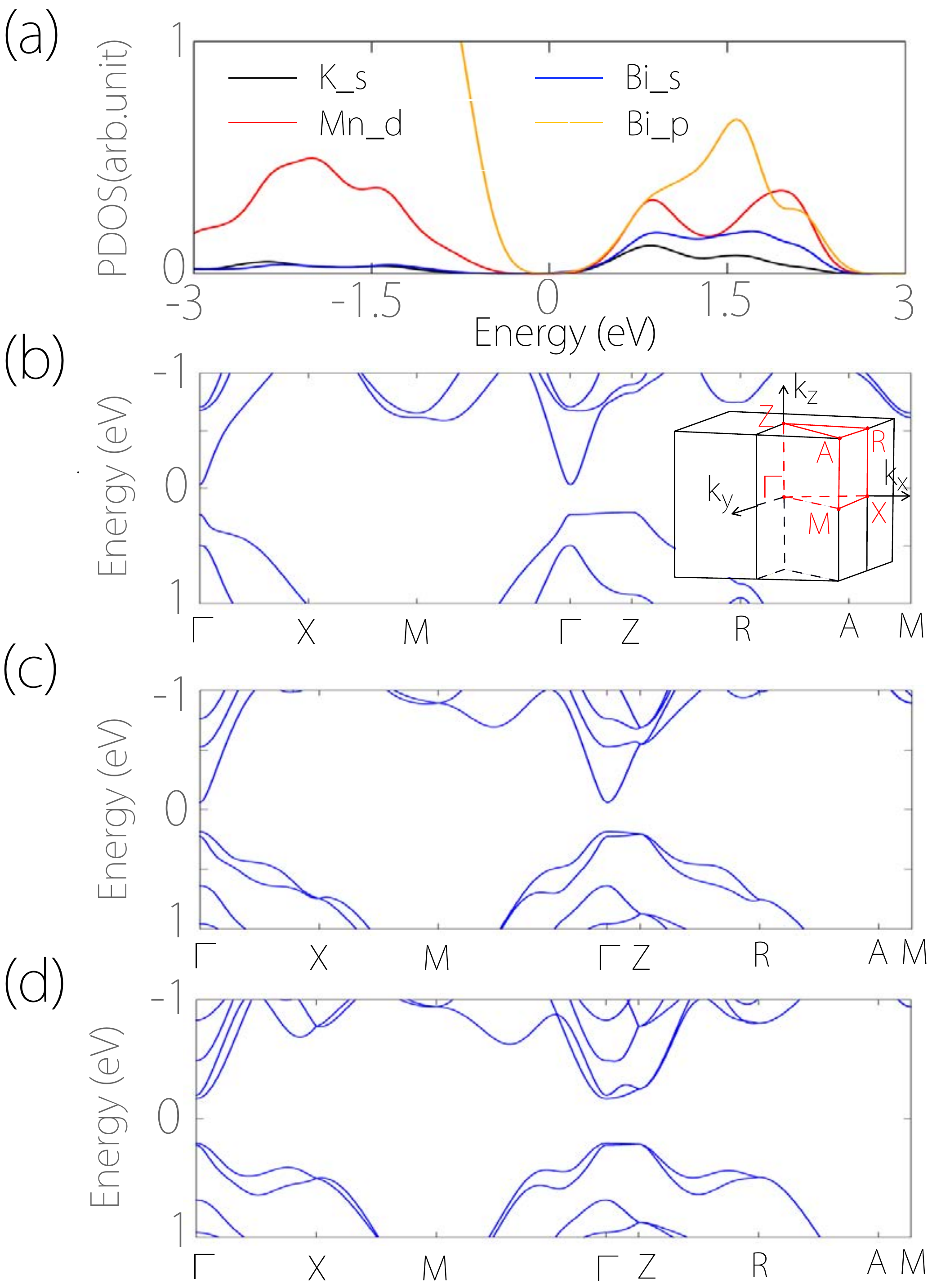}}
	\caption{\label{fig:AFM_bs} (a) Orbital projected density of states (PDOS) for KMnBi. (b-d) Calculated band structures for (b) KMnBi, (c) RbMnBi, and (d) CsMnBi. The inset in (b) shows the Brillouin zone. SOC is included in the calculation.}
\end{figure}

After determining the magnetic ordering, we now turn to the electronic properties. In Fig.~\ref{fig:AFM_bs}, we plot the calculated electronic band structures and the orbital projected density of states (PDOS). (Here we only show PDOS for KMnBi, as the essential features are similar for the other two. SOC is included in the calculation.) One clearly observes that these materials are narrow-gap semiconductors, with gap values of 0.26 eV, 0.25 eV, and 0.36 eV for KMnBi, CsMnBi, and RbMnBi, respectively. One notes that RbMnBi and CsMnBi have direct gaps at the $\Gamma$ point, whereas KMnBi is indirect-gap: its conduction band minimum (CBM) is located at $\Gamma$, but its  valence band maximum (VBM) is at $Z$. The low-energy bands are dominated by Bi-$p$ and Mn-$d$ orbitals. In addition, in Fig.~\ref{fig:AFM_bs}, each band is doubly degenerate. This is due to the existence of the $\mathcal{PT}$ symmetry. Interestingly, for both C-type and G-type AFM, although $\mathcal{T}$ and $\mathcal{P}$ are separately broken, their combination ($\mathcal{PT}$) is still preserved.

More interestingly, one observes a salient difference in the band dispersion between the conduction band minimum (CBM) and the valence band maximum (VBM), especially for KMnBi and RbMnBi. Around CBM, the conduction band has strong dispersion both in-plane ($\Gamma$-$X$ and $\Gamma$-$M$) and out-of-plane ($\Gamma$-$Z$). In contrast, around VBM, the dispersion is much suppressed, especially along the out-of-plane direction. This difference can be quantitatively captured by the effective masses. As shown in Table~\ref{table:cm}, the electron effective masses for these materials are quite small (except for $m_z^*$ in CsMnBi).
In contrast, for most cases, the hole effective mass is at least an order of magnitude larger. For example, in KMnBi, $m_z^{e*}\approx 0.011 m_0$ ($m_0$ is the free electron mass); meanwhile, $m_z^{h*}\approx 2.296 m_0$, which is over 200 times larger.

\begin{figure}[!t]
	{\includegraphics[clip,width=7.4cm]{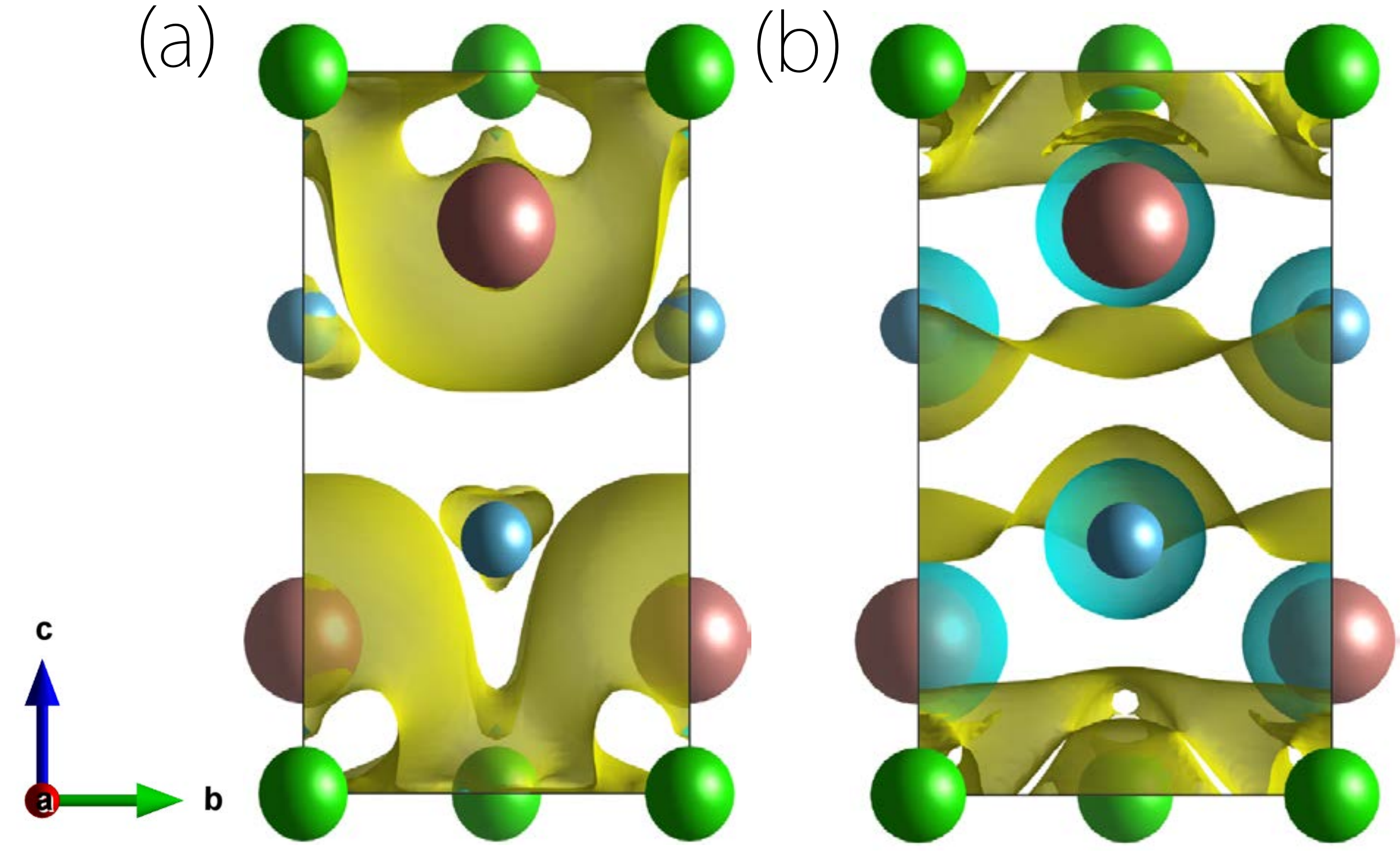}}
	\caption{\label{fig:cd} Charge distribution (side view of the unit cell) for (a) VBM state and (b) CBM state of KMnBi. Here the isosurface value of 0.0015 eV/\AA$^3$ is taken. }
\end{figure}

To understand this difference, we plot the charge distribution for the two states at CBM and VBM of KMnBi, as shown in Fig.~\ref{fig:cd}. One can see that the CBM state is more extended in the out-of-plane direction, whereas the VBM state is mostly confined within each quintuple layer. This is consistent with the observation of the much larger dispersion for the conduction band.

\begin{table}
	\centering
	\caption{Calculated effective masses and carrier mobilities for the three materials. Here, $m_0$ is the free electron mass. The mobilities are evaluated at 300 K and are shown in unit of 10$^{3}$ cm$^{2}$/(V$\cdot$s).}
	\label{my-label}
	\renewcommand\arraystretch{1.4}
	\begin{tabular}{p{1.0cm}<{\centering}p{1.75cm}<{\centering}p{1.25cm}<{\centering}p{1.25cm}<{\centering}p{1.25cm}<{\centering}p{1.25cm}<{\centering}p{1.4cm}<{\centering}}
		\hline\hline
		\rule{0pt}{13pt}
		 &Carrier type   &$m_{x}^{*}/m_{0}$   &$m_{z}^{*}/m_{0}$
	    &$\mu_{x}^\text{3D}$  &$\mu_{z}^\text{3D}$ \\
		%&$ $   &$ $ &$ $         & \multicolumn{2}{c}{(10$^{3}$ cm$^{2}$/(V$\cdot$s))}\\
		\hline
		
		\multirow{2}*{KMnBi} &e  &$0.040$  &$0.011$   &$73.4$  &$189$  \\
		 &h   &$0.281$  &$2.296$   &$0.979$  &$0.244$  \\
		\hline
		
		\multirow{2}*{RbMnBi} &e   &$0.062$  &$0.138$   &$7.34$  &$1.77$  \\
	   &h  &$0.311$  &$5.741$    &$0.0797$  &$0.0052$  \\
		\hline
		
			\multirow{2}*{CsMnBi} &e   &$0.257$  &$1.311$   &$0.161$  &$0.211$  \\
		 &h  &$0.343$  &$13.209$   &$0.0637$  &$0.0031$  \\
		\hline\hline
	\end{tabular}
	\renewcommand\arraystretch{1.4}
	\label{table:cm}
\end{table}

This distinct feature in band dispersion will directly manifest in the carrier transport properties. Here, we estimate the carrier mobilities by using the deformation potential theory~\cite{bardeen1950deformation}. With effective mass approximation, the intrinsic mobility for transport along the $i$ direction can be obtained from the formula~\cite{bardeen1950deformation,xi2012first}
\begin{equation}
\mu^\text{3D}_i=\frac{2\sqrt{2\pi}e \hbar^{4} C^\text{3D}_i}{3(k_{B}T)^{3/2}(m^{*}_{d})^{3/2}m_i^{*}(\mathcal{D}_i)^{2}}.
\label{eq:3D_mobility}
\end{equation}
Here, $C^\text{3D}_i=(1/V_0)(\partial^{2} E_S/\partial \varepsilon_i^{2})$
is the 3D elastic constant, $E_S$ and $V_0$ are the energy and the volume of the system, $\varepsilon_i=(\ell_i-\ell_i^0)/\ell_i^0$ is the strain along the $i$ direction, $m^{*}_{d}=(m^{*}_{x}m^{*}_{y}m^{*}_{z})^{1/3}$ is the average effective mass, and $\mathcal{D}_i=\partial \Delta/\partial \varepsilon_i$ is the deformation potential constant, with $\Delta$ the shift of the band edge energy under strain. We evaluate the mobility at room temperature ($T=300$ K). The obtained results for electron and hole carriers are listed in Table~\ref{table:cm}.

One observes that the mobilities for electron carriers are quite high. Remarkably, for KMnBi, it can even reach the order of $10^5$ cm$^2$/(V$\cdot$ s) at 300 K. This value is much higher than the crystalline Si ($\sim 1400$ cm$^2$/(V$\cdot$ s)), and is comparable to the carbon nanotubes~\cite{durkop2004extraordinary}. In comparison, the hole mobilities here are much lower. In KMnBi, the hole mobility along the $c$-axis is about 800 times lower than the electron mobility. The high electron mobility and the asymmetry between electron and hole transport properties could be useful for AFM spintronics applications.

\section{2D Single Layer}

Since the $A$MnBi family materials have a layered structure, it is possible to fabricate ultrathin layers of these materials, e.g., by exfoliation method or by bottom-up growth method. In the following, we shall investigate the properties of the 2D single layer (SL) form of the $A$MnBi family materials. The three materials in 2D exhibit similar features, so in the discussion, we will focus on the results of SL-KMnBi. The results for the other two are presented in Tabel~\ref{my-label} and shown in the Supplemental Material~\cite{supp_material}.

\begin{table*}[!t]
	\centering
	\caption{Calculated structural, magnetic, and transport properties for SL-$A$MnBi materials. $a$ is the lattice constant. The $E$'s are the energy per formula unit for the different magnetic configurations. For example,
${E}_\text{FM}^x$ denotes the energy for the FM state with spins along the $x$ direction. The ground state energy is taken as the reference (zero point). The mobilities are evaluated at 300 K. }
	\label{my-label}
	\renewcommand\arraystretch{1.4}
	
	\begin{tabular}{p{2.0cm}<{\centering}p{1.0cm}<{\centering}p{1.0cm}<{\centering}p{1.5cm}<{\centering}p{1.5cm}<{\centering}p{1.5cm}<{\centering}p{1.25cm}<{\centering}p{1.5cm}<{\centering}p{1.5cm}<{\centering}p{1.5cm}<{\centering}}

		\hline\hline
		\rule{0pt}{13pt}
	
		&$a$ & ${E}_\text{FM}^x$   &${E}_\text{AFM}^x$  &${E}_\text{FM}^z$  &${E}_\text{AFM}^z$  &${T}_\text{N}$  &$\mu_{x}^\text{2D}(e)$ &$\mu_{x}^\text{2D}(h)$ \\
&(\AA) &(meV)  &(meV) &(meV)  &(meV)   &(K)
 &\multicolumn{2}{c}{(10$^{3}$ cm$^{2}$/(V$\cdot$s))}  \\
		\hline
		
		KMnBi  &$4.802$    &$72.32$  &$1.10$  &$68.59$  &$0$  &$305$  &$2.72$
		&$2.59$\\
	
		\hline
		
		RbMnBi  &$4.845$    &$75.86$   &$1.01$  &$71.97$  &$0$  &$307$  &$4.00$  &$3.74$\\
		\hline
		
		CsMnBi   &$4.943$    &$64.98$   &$0.86$  &$60.96$  &$0$  &$302$  &$4.70$ &$6.09$ \\
		\hline\hline
	\end{tabular}
	\renewcommand\arraystretch{1.4}
	\label{table:2D}
\end{table*}

\subsection{Structure and Magnetic Ordering}

\begin{figure}[!t]
	{\includegraphics[clip,width=8.2cm]{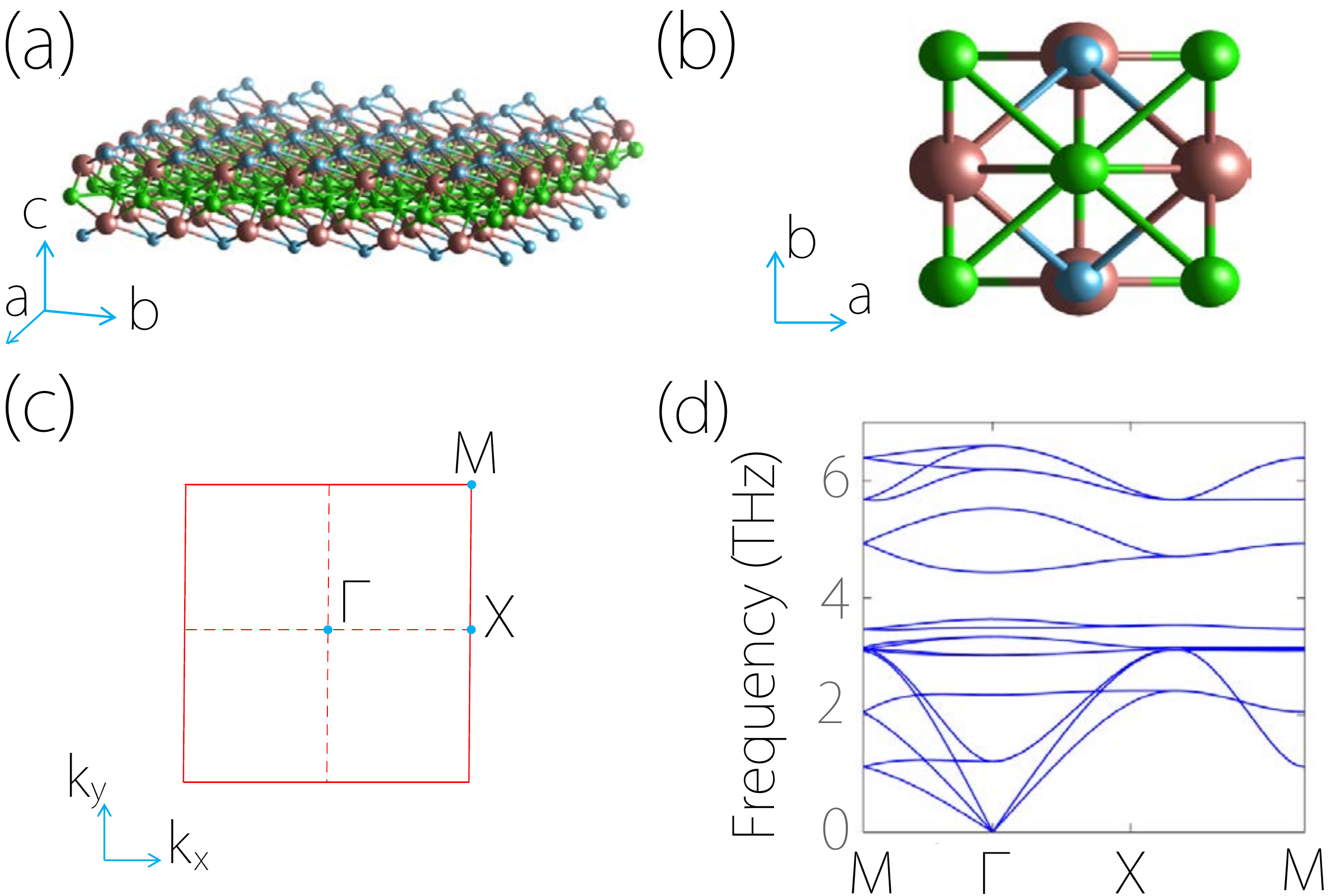}}
	\caption{\label{fig:2D_stru} (a) Perspective view and (b) top view of SL-$A$MnBi. (c) 2D BZ with the three high-symmetry points labeled. (d) Phonon dispersion of SL-KMnBi. }
\end{figure}

Figure~\ref{fig:2D_stru}(a) and \ref{fig:2D_stru}(b) show the optimized single layer structure. The calculated lattice parameters are shown in Table~\ref{table:2D}. Compared to the 3D bulk, one observes that the lattice constants in the $a$-$b$ plane are slightly larger.

To check the structural stability, we have calculated the phonon spectrum. The result for SL-KMnBi is shown in Fig.~\ref{fig:2D_stru}(d). One can see that there is no imaginary frequency (soft mode) in the spectrum, indicating that the 2D structure is dynamically stable.

In Sec.~\ref{3D_Bulk}, we have found that in the 3D bulk, the $A$MnBi family materials have strong AFM coupling within the Mn layer (due to the strong superexchange interaction), while the interlayer coupling is relatively small. Hence, one naturally expects that in the SL structure, the AFM ordering should still be preferred. This is indeed the case. In Table~\ref{table:2D}, we have compared the energies for the FM and AFM configurations. One finds that AFM is favored for all three materials in the ground state. And the magnetic easy axis remains to be along the $z$ direction.
The Neel temperatures are estimated by the Monte-Carlo simulations. The results are listed in Table~\ref{table:2D}. One can see that even in the 2D limit, the transition temperatures are still above 300 K, much higher than that for the 2D FePS$_3$ ($\sim 118$ K).

\subsection{Electronic Property and Mobility}

In Fig.~\ref{fig:strain_neel}(a), we plot the calculated band structure for SL-KMnBi. One observes that the material is a narrow gap semiconductor. Interestingly, in contrast to its 3D bulk, the 2D SL has a direct band gap ($\sim$ 0.15 eV) at the $\Gamma$ point. In addition, the bands show strong dispersion around both CBM and VBM. The calcualted effective masses are $m^{e*}=0.084 m_{0}$ and  $m^{h*}=0.082 m_{0}$.

In 2D, the carrier mobilities can be estimated by using the following formula~\cite{bardeen1950deformation,xi2012first}
\begin{equation}
\mu^\text{2D}_i=\frac{e\hbar^{3}C^\text{2D}_i}{k_{B}T m^{*}_{d}m^{*}_i(\mathcal{D}_i)^{2}},
\label{eq:mobility}
\end{equation}
where $C^\text{2D}_i=(1/S_0)(\partial^{2} E_S/\partial \varepsilon_i^{2})$
is the 2D elastic constant, $m^{*}_{d}=(m^{*}_{x}m^{*}_{y})^{1/2}$, and the other symbols carry the same meaning as in Eq.~(\ref{eq:3D_mobility}). The obtained values (for $T=300$ K) are presented in Table~\ref{table:2D}. One finds that the mobilities for both electron and hole carriers are on the order of $10^3$ cm$^2$/(V$\cdot$ s). Although this value is lower than that of graphene [$\sim$ 120,000 cm$^2$/(V$\cdot$s) at 240 K]~\cite{bolotin2008temperature}, it is comparable to that of phosphorene [10$^3$ cm$^2$/(V$\cdot$s) at 300 K]~\cite{li2014black} and much higher than that of 2D MoS$_2$ [$\sim$ 200 cm$^2$/(V$\cdot$ s) at room temperature]~\cite{radisavljevic2011single}.

\subsection{Magnetic and Metal-Insulator Phase Transition}

\begin{figure}[!t]
	{\includegraphics[clip,width=8.2cm]{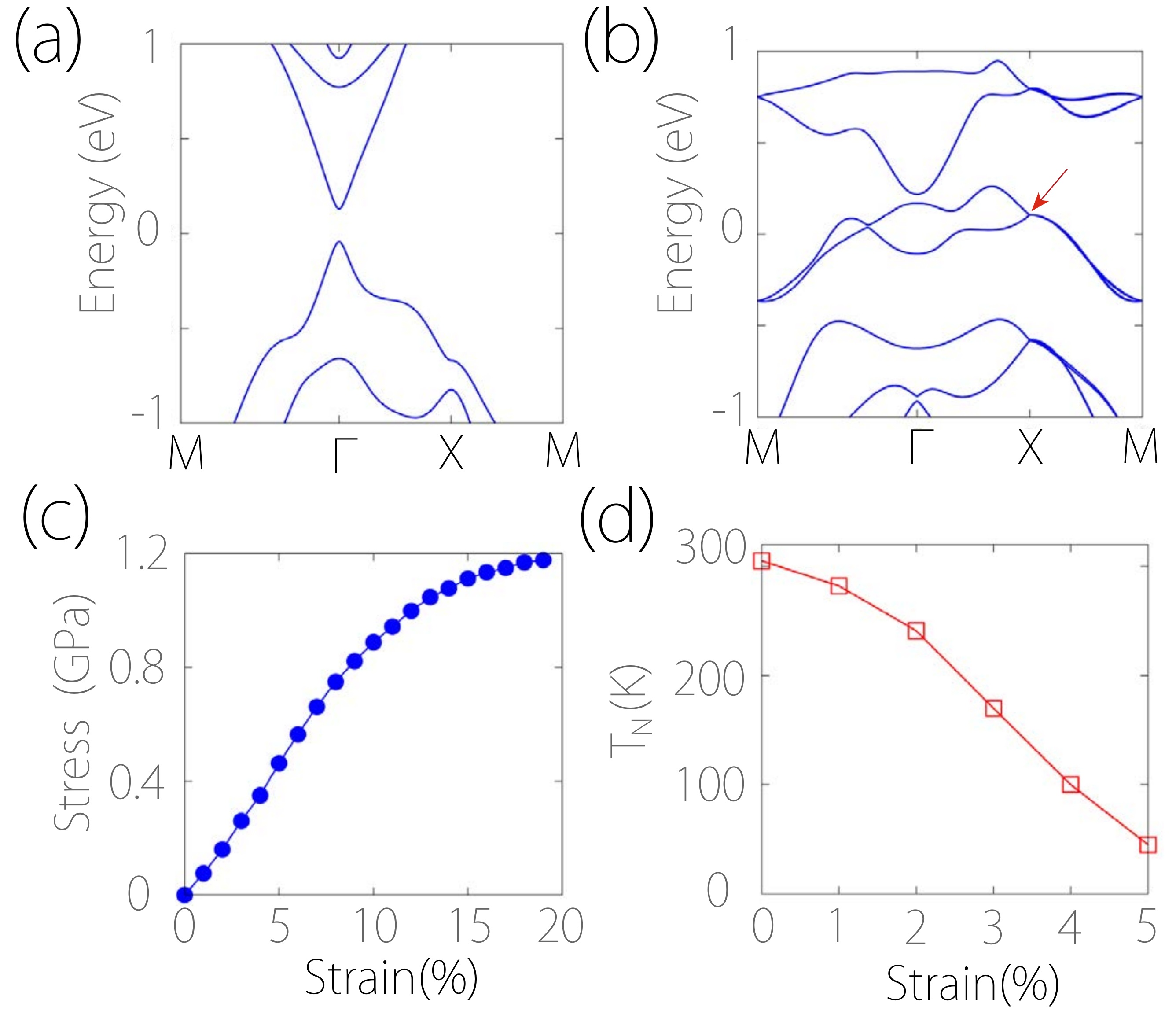}}
	\caption{\label{fig:strain_neel} Calculated band structures for SL-KMnBi (a) in the AFM ground state, and (b) in the paramagnetic state. The SOC is included in these results. (c) Stress-strain relation for SL-KMnBi. (d) Variation of the Neel temperature versus the applied strain for SL-KMnBi.}
\end{figure}

Since the magnetic transition temperature for SL-KMnBi is around the room temperature, the phase transition can be readily achieved and studied in experiment. We have also investigated the paramagnetic phase above the phase transition. Its band structure is shown in Fig.~\ref{fig:strain_neel}(b). One observes that this phase is a metal. Therefore, the magnetic phase transition is simultaneously also a metal-insulator phase transition. This dual character of the transition may make the material a promising candidate for functional devices.

In addition, we observe that close to the Fermi level, the band structure in Fig.~\ref{fig:strain_neel}(b) has a degeneracy at the $X$ point (indicated by the red arrow). As each band here has a double degeneracy due to the $\mathcal{PT}$ symmetry, this point is  a fourfold degenerate Dirac point. Importantly, note that this Dirac point is stable with SOC fully considered, distinct from the case in graphene (where the SOC in principle gaps the Dirac point). It belongs to the so-called 2D spin-orbit Dirac point, first proposed by Young and Kane~\cite{young2015dirac}. The 2D spin-orbit Dirac point was previously predicted in monolayer HfGeTe family materials~\cite{guan2017two}, and more recently experimentally confirmed in $\alpha$-Bismuthene~\cite{kowalczyk2020realization}. For SL-KMnBi, we note that its symmetry in the paramagnetic phase is identical to that of SL-HfGeTe studied in Ref.~\cite{guan2017two}, thus the Dirac point at $X$ also share the same symmetry protection, namely, it is protected by $\mathcal{P}$, $\mathcal{T}$, and the nonsymmorphic glide mirror $G_z:(x,y,z)\rightarrow (x+1/2,y+1/2,-z)$. The detailed symmetry analysis and the effective model can be found in Ref.~\cite{guan2017two}.

One crucial advantage of 2D materials is that their properties can be readily tuned by external means. Particularly, 2D materials can usually sustain large strains~\cite{lee2008measurement,kim2009large}. Here, we consider the effect of lattice strain on the phase transition. In Fig.~\ref{fig:strain_neel}(c), we plot the calculated strain-stress relation for SL-KMnBi. One finds that the material can sustain a linear elastic regime up to $\sim7$\% strain, and the critical strain is beyond 20\%.  For strains within the linear elastic regime, we repeat the Monte-Carlo simulations to investigate the change in the magnetic transition temperature. The result is shown in Fig.~\ref{fig:strain_neel}(d). One observes that the magnetic transition temperature (hence the magnetic ordering) is strongly suppressed by the applied strain. For example, a 5\% strain can lower the transition temperature by more than 250~K.

This sensitive dependence of magnetism on strain may open many interesting possibilities for applications. For example, at a fixed temperature (close to room temperature), a slight applied strain can suppress the magnetic ordering and make a transition from insulator to metal. This could be useful for novel information storage devices and for sensitive pressure/stress sensors.

%\begin{figure}[!t]
%{\includegraphics[clip,width=8.2cm]{WP_FM.pdf}}
%\caption{\label{fig:WP_FM} Bulk electronic structures of (a) KMnBi and (b) CsMnBi with FM configuration as SOC is introduced. The energy position of two pairs of WPs are at $E_{f}-0.0012$ eV and $E_{f}-0.0016$ eV, respectively. (c) The evolution of WCC in the plane of $k_{z}=0$. The distribution of Berry curvature around WP with topological charge (d) $C=1$ and (e) $C=-1$. (f) Fermi surface on the (100) surface at the nodal energy. }
%\end{figure}

%%\begin{figure}[htbp]
%{\includegraphics[clip,width=8.2cm]{strain_KMnBi.pdf}}
%\caption{\label{fig:strain_KMnBi} (a) Bulk  band structures in A-type AFM KMnBi material (included SOC) and (b) enlarged low-energy band along the $\Gamma$-Z directin, showing the small hybrid band gap ($\sim$ 10meV) around the crossing. (c) The corresponding WCC evolution on the $k_{z}=0$ plane and (d) the projected spectrum for the (100) surface.}
%\end{figure}

\section{Discussion and Conclusion}

We have a few points to be discussed before closing. First, for the 3D bulk \emph{A}MnBi family, their single crystals have been synthesized in experiment. The magnetic orderings obtained here are consistent with the previous neutron scattering experiment~\cite{Schucht1999ChemInform}. The electronic and transport properties of these materials have not been reported before. Here, we predict that these materials can have very high mobility. For KMnBi, it even reaches $10^5$ cm$^2$/(V$\cdot$s) for electron carriers. Such high-mobility AFM materials are of great interest from application perspectives.

Second, the 2D SLs of these \emph{A}MnBi materials are found to be stable. They may be obtained by exfoliation method from the bulk, or by the bottom-up growth method such as the molecular beam epitaxy. Regarding the exfoliation method, we have estimated that the exfoliation energy for SL-KMnBi is about 0.71 J/m$^{2}$, which is comparable to that of graphene
($\sim$0.37 J/m$^2$)~\cite{zacharia2004interlayer} and MoS$_2$ ($\sim$0.41 J/m$^2$)~\cite{zhao2014obtaining}, and is less than that
of Ca$_2$N ($\sim$1.14 J/m$^2$)~\cite{guan2015electronic}.

\begin{figure}[htbp]
{\includegraphics[clip,width=8.2cm]{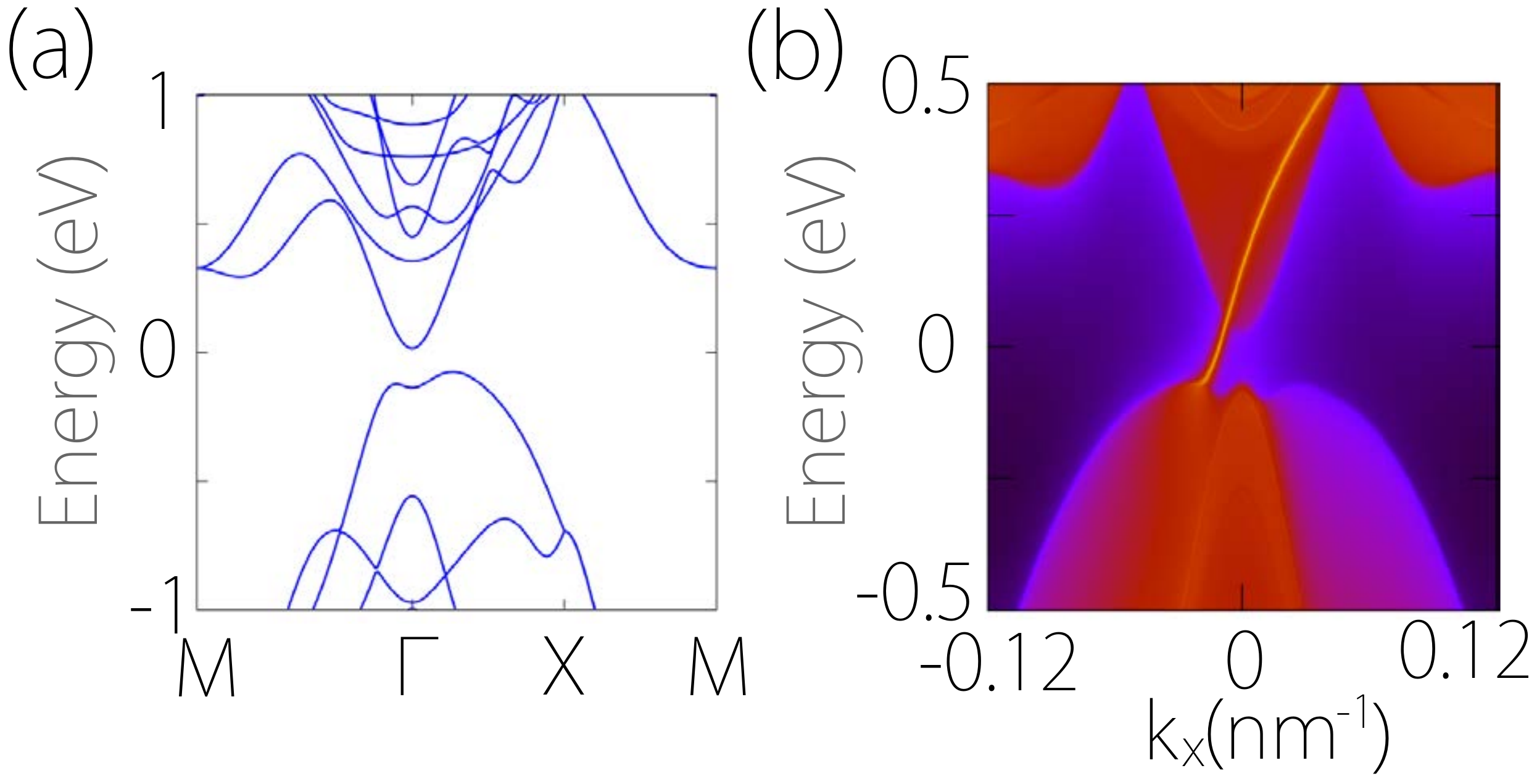}}
\caption{
(a) Band structures of  SL-KMnBi in the FM state. (b) shows the corresponding edge spectrum, in which one observes a gapless chiral edge band.}\label{fig:QAH}
\end{figure}
Third, because properties of 2D materials can be more readily tuned, it is possible to control the magnetic ordering
in SL-\emph{A}MnBi, e.g., by proximity coupling to a magnetic substrate or by applied magnetic fields [applied strain can further promote this possibility, as shown in Fig.~\ref{fig:strain_neel}(d)].  We find that if the magnetic ordering in SL-KMnBi can be tuned into FM, then the resulting state will be a quantum anomalous Hall (QAH) insulator. As shown in Fig.~\ref{fig:QAH}(a), the band structure for the FM state remains a semiconductor with a gap $\sim 100$ meV. By analyzing the evolution of Wannier charge centers, we confirm that the system has a nontrivial Chern number of 1. This indicates that at the boundary of the system, there must exist one gapless chiral edge band, which is confirmed by the calculated edge spectrum in Fig.~\ref{fig:QAH}(b). In addition, we also mention that if FM ordering can be realized in bulk KMnBi, the resulting state will be a topological metal with type-II Weyl points~\cite{supp_material}.

In conclusion, we have systematically investigated the physical properties of the $A$MnBi($A=$K, Rb, Cs)-family materials. We show that these materials are room-temperature AFM narrow-gap semiconductors. The calculated magnetic configurations are consistent with the previous neutron diffraction experiment. Remarkably, we find very high electron mobilities for these materials, which can even reach $10^5$ cm$^2$/(V$\cdot$s) in the case of KMnBi. In contrast, the hole mobilities are much lower. This feature permits a possibility to control the transport via different types of doping. In the 2D single layer form, these materials maintain robust AFM ordering. The Neel temperatures ($\sim 300$ K) are much higher than the existing 2D AFM materials. The mobilities for these single layers are still fairly high ($\sim 10^3$ cm$^2$/(V$\cdot$s)). Interestingly, the magnetic phase transition in the single layer is also a metal-insulator phase transition, with the paramagnetic metal phase possessing a pair of 2D spin-orbit Dirac points protected by the nonsymmorphic space group symmetry.
We find that magnetism can be effectively controlled by strain. For SL-KMnBi, a 5\% strain can decrease the Neel temperature by more than 250 K. We further show that if the magnetic ordering can be turned into FM, the system can become a QAH insulator with gapless chiral edge states. Our work reveals a range of fascinating properties for the $A$MnBi($A=$K, Rb, Cs)-family materials. Besides AFM spintronics, the interplay between magnetism, high-mobility, and lattice strain may lead to applications of these materials in many possible novel functional devices.

\begin{acknowledgements}
The authors thank Jingsi Qiao, Shuai Dong, and D. L. Deng for valuable discussions. This work is supported by the National Natural Science Foundation of China (NSFC) (Grant No.~11704117 and 11974076), the Singapore Ministry of Education Academic Research Fund Tier 2 (MOE2019-T2-1-001) and Natural Science Foundation of Fujian Province of China (Grant No. 2018J06001).
We acknowledge computational support from Texas Advanced Computing Center and H2 clusters in Xi'an Jiaotong University.
\end{acknowledgements}

%\appendix

\bibliographystyle{apsrev4-1}
\bibliography{FWS_refs}

\end{document}